\documentclass[twocolumn]{autart}
\usepackage{}
\usepackage{amsfonts}    

\usepackage{amsmath}
\usepackage{amssymb}
\usepackage{graphicx,color}          
\usepackage[dvips]{epsfig}    

\begin{document}

\begin{frontmatter}

\title{Analysis and Control of Quantum Finite-level Systems Driven by Single-photon Input States\thanksref{footnoteinfo}} 

\thanks[footnoteinfo]{This paper was not presented at any IFAC
meeting. Corresponding author Y.~Pan.}

\author[Yu]{Yu Pan}\ead{yu.pan.83.yp@gmail.com},    
\author[Guofeng]{Guofeng Zhang}\ead{Guofeng.Zhang@polyu.edu.hk},               
\author[Matt]{Matthew~R.~James}\ead{matthew.james@anu.edu.au}  

\address[Yu]{Research School of Engineering, Australian National University, Canberra, ACT 0200, Australia}  
\address[Guofeng]{Department of Applied Mathematics, The Hong Kong Polytechnic University, Hong Kong, China}             
\address[Matt]{ARC Centre for Quantum Computation and Communication Technology, Research School of Engineering, Australian National University,
Canberra, ACT 0200, Australia}        


\begin{abstract}                          
Single-photon states, which carry quantum information and coherently interact with quantum systems, are vital to the realization of all-optical engineered quantum networks. In this paper we derive the analytical form of the output field state for a large class of quantum finite-level systems driven by  single-photon input field states using a transfer function approach. Single-photon pulse shaping via coherent feedback is also studied.
\end{abstract}

\end{frontmatter}

\section{Introduction}
In classical (non-quantum) control theory, responses of systems to various types of input signals reveal important system properties. For example, in the linear case, step response tells us the rise time, overshoot and settling time of the system, frequency response shows the ability of the system to track rapidly changing signals, impulse response allows to calculate the $H_2$ norm of the system, response to $L_2$ signals reveals the robustness of the system to disturbance measured in terms of $H_\infty$ performance, and response to Gaussian white noise is the foundation of the celebrated Kalman filtering theory, which is the basis of Linear-Quadratic-Gaussian (LQG) feedback control. In the quantum regime, the response of quantum systems to quantum Gaussian input states has been studied intensively due to the prevalent use of Gaussian states such as vacuum states, coherent states and squeezed states. This is the basis of widely applied measurement-based quantum feedback control in quantum optics \cite{WM10}.

However, besides Gaussian states,  there are many other useful quantum states such as single-photon and multi-photon states. Simply speaking, a light field is in an $n$-photon state if it contains exactly $n$ photons. When $n=1$, the light field is in a single-photon state. When $n\geq 2$, we simply say it is in a multi-photon state. Single-photon states and multi-photon states are very useful resources for quantum information technology. For example, photons are ideal information carriers that transfer quantum information from one node to another node in a quantum network \cite{Cirac97,Nysteen14,Nielsen04}.  It is possible to build a quantum switch \cite{Chen13} which uses a single photon as controller to switch on and off a physical process. last but not the least, the ability to control the flow of single photons using on-chip finite-level systems could give birth to a new generation of light transistors \cite{Chang07}.  Therefore, the analysis and control of quantum systems driven by single-photon or multi-photon states is important for a successful quantum engineering.

In the linear regime, the response of quantum systems to single-photon and multi-photon states has been recently studied in \cite{Guofeng13,Guofeng14}. Moreover, interestingly, it is shown that linear quantum systems theory turns out quite useful in the study of quantum memories where a single-photon is efficiently stored and read out by a collection of atoms \cite{Hush13,Yamamoto14,Nurdin14}.

In this paper we go beyond the linear regime and study quantum finite-level systems driven by single-photon states. Quantum finite-level systems,  for example two-level atoms, are {\it nonlinear} quantum systems. The study of the interaction between finite-level systems and single-photon states are fundamental to the study of light-matter interaction, which is the foundation of engineered integrated quantum networks \cite{Cirac97}. The interaction between finite-level systems and single-photon states has been studied extensively in the quantum optics community \cite{Cirac97,Shen05,Zhou08,Fan10,Gough12,Naoki14}. Unfortunately, most studies are usually based on various  assumptions such as weak excitation limit and primarily focused on two-level systems. In this paper we show from a control theoretic point perspective that, for a large class of quantum finite-level systems driven by single-photon states, the analytical expression of the output field state can be derived, cf. Theorem \ref{theorem3}. Interestingly, due to the special nature of single-photon states, the techniques developed for the linear case in \cite{Guofeng13,Guofeng14} can be adopted here to show that the pulse shape of the output single photon is obtained via the linear transfer of the input single-photon pulse shape. Based on this analysis result, we  also investigate how to manipulate the pulse shape of single-photon wavepackets by means of coherent feedback.


The rest of the paper is organized as follows. In Section \ref{secss}, we introduce the existing results on quantum stochastic differential equations, the response of linear systems to single-photon states, and finite-level systems (Subsection \ref{subsec:finite_level_single_photon}). In Section \ref{secfl}, we prove the main result that the input-output relation for a large class of finite-level systems can be solved using a transfer function approach. In Section \ref{secapp}, we present some applications of the main result. We study in Section \ref{sec:control} how to use coherent feedback to manipulate the pulse shape of single-photon wavepackets. Conclusion is put in Section \ref{seccon}.

\section{Notations and preliminaries}\label{secss}
We use $X^\dagger$ to denote the adjoint of an operator $X$ defined on a Hilbert space $\mathcal H$. The notation $\ast$ stands for complex conjugation, and $T$ for the transpose. $X^\dagger=X^\ast$ if $X$ is a one-dimensional scalar. The commutator of two operators is given by $[A,B]=AB-BA$. We also define the doubled-up column vector of operators as $\breve{X}=[X^T,X^\dagger]^T$. We use $\delta(\cdot)$ for the Dirac-delta function, and the symbol $\otimes$ for the Kronecker tensor product. $\Re(\cdot)$ and $\Im(\cdot)$ are the real part and imaginary part of a number. The Fourier transform of a function $f(t)$ is defined by $\tilde{f}(\omega)= \mathcal F[f](\omega) :=\int_{-\infty}^\infty e^{-\mbox{i}\omega t}f(t)dt$.

\subsection{Open quantum systems}
An open quantum system often involves a plant interacting with external environment which is defined on a Fock space $\mathcal{H}_B$
over $L^2 (\mathbb{R}_+ , dt) $.   The open quantum system can be properly modelled using a triplet $(S,L,H_0)$ \cite{hudson84,Gough09,Gough12}. $S$ is a constant scattering matrix. The plant is coupled to the external fields through the operator $L$. We assume $L=[c_1L_0\ \cdot\cdot\cdot\ c_KL_0]^T=\theta^TL_0$, that is, the plant couples to the environment through $K$ channels via the same coupling operator $L_0$. $H_0$ is the Hamiltonian of the plant. The state of the total system (plant plus field) undergoes a unitary evolution generated by a unitary operator $U(t,t_0)$ whose dynamics is given by
\begin{equation}\label{lem1p}
dU(t,t_0)=\{b^\dag(t)L-L^\dag Sb(t)-(\frac{1}{2}L^\dag L+\mbox{i}H_0)dt\}U(t,t_0)
\end{equation}
for $t\geq t_0$, where $t_0$ is the initial time. In Heisenberg picture, the evolution of a plant operator $X$ is given by $X(t)=U^\dagger(t,t_0)(X\otimes I)U(t,t_0)$, where $I$ is the identity operator on $\mathcal H_B$. Driven by canonical input fields, the dynamics of $X(t)$ is described by quantum stochastic differential equations (QSDEs) of the form
\begin{eqnarray}\label{nopr2}
\dot{X}(t)&=&\mathcal G_t(X)+b^\dagger(t)S^\dagger[X(t),L(t)]\nonumber\\
&+&[L^\dagger(t),X(t)]Sb(t),\\
b_{out}(t)&=&L(t)+Sb(t),\nonumber
\end{eqnarray}
where the generator $\mathcal G_t(X)$ is defined as
\begin{eqnarray}\label{pn1}
\mathcal G_t(X)&:=&-\mbox{i}[X(t),H_0(t)]+\sum_{k=1}^K(L_k^\dagger(t) X(t)L_k(t)\nonumber\\
&&-\frac{1}{2}L_k(t)^\dagger L_k(t)X(t)-\frac{1}{2}X(t)L_k^\dagger(t) L_k(t)).
\end{eqnarray}
By the form of the operator $L$ discussed above, we have $L_k=c_kL_0$ for $k=1,...,K$. In Eq. (\ref{nopr2}), $b(t)=[b_1(t)\ \cdots \ b_K(t)]^T$ is a vector of annihilation operators for input field modes which satisfy the canonical commutation relation $[b_i(t),b_j^\dagger(s)]=\delta(t-s),\ i=j$ and $[b_i(t),b_j^\dagger(s)]=0,\ i\neq j$ resembling classical white noise if the field is vacuum. Physically, $b_i(t)$ and $b_i^\dag(t)$ can be understood as the annihilation and creation of one photon in the $i$-th channel at time $t$.  $b_{out}(t)=U^\dagger(t,t_0)b(t)U(t,t_0)$ defines the annihilation operators of the output fields. For later use, we define $f(t,t_0)=[f_1(t,t_0)\ \cdot\cdot\cdot\ f_K(t,t_0)]^T := U(t,t_0)b(t)U^\dagger(t,t_0)$.  Finally, if $\rho$ is the state of the total system, the state of the field can be obtained by tracing out the plant \cite{Nielsen04}, that is, $\rho_{field} =\text{Tr}_s(\rho):=\sum_j\langle j_s|\rho|j_s\rangle$, where $\{|j_s\rangle\}$ is the basis of system Hilbert space $\mathcal H$.

\subsection{Response of linear quantum systems to single photon input}
If the plant is a collection of quantum harmonic oscillators, Eq. (\ref{nopr2}) describes a linear quantum system as
\begin{eqnarray}\label{lintem}
\dot{\breve{x}}(t)&=&A\breve{x}(t)+B\breve{b}(t),\nonumber\\
\breve{b}_{out}(t)&=&C\breve{x}(t)+D\breve{b}(t),
\end{eqnarray}
where the constant matrices $A,B,C,D$ can be expressed by $S,L,H_0$, see e.g. \cite{Guofeng13,Guofeng14}. The input-output relation of this system can be written as
\begin{equation}\label{outlin}
\breve{b}_{out}(t)=Ce^{A(t-t_0)}\breve{x}(0)+\int_{t_0}^t g_{G}(t-r)\breve{b}(r)dr,
\end{equation}
where
\begin{equation} \label{tf}
g_{G}(t):=\left\{
\begin{array}{ll}
\delta(t)D+Ce^{At}B, & t\geq 0, \\
0, & t<0.
\end{array}
\right.
\end{equation}
The response of quantum linear systems to single photon input has been studied in detail in \cite{Guofeng13,Guofeng14}. A single-channel single photon input is defined by, \cite{Milburn08,Gough12,Guofeng13,Guofeng14},
\begin{equation}\label{sinput}
|1_\xi\rangle = \int_{-\infty}^\infty \xi(r)b^\dagger(r)|0\rangle dr,
\end{equation}
where $\xi(r)$ represents the pulse shape of the single photon in the time domain. Here $|0\rangle$ denotes a one-channel vacuum input. $|\xi(r)|^2dr$ is the probability of finding the photon in the time interval $[r,r+dr)$, and we have a normalization condition $\int_{-\infty}^\infty|\xi(r)|^2dr=1$.  Denote by $G(s)$  the transfer function determined by $g_{G}(t)$ in Eq. (\ref{tf}). Then it is shown in \cite{Guofeng13} that the pulse shape $\xi^{'}(t)$ of the output field state is given in terms of the input-output transfer $\tilde{\xi}^{'}(\omega)=G(\mbox{i}\omega)\tilde{\xi}(\omega)$. For multiple-channel input, the single photon state may be defined as a superposition of one-photon excitation on multiple channels and thus given by
\begin{equation}\label{sinput1}
|1_\xi\rangle=\sum_{k=1}^K\int_{-\infty}^\infty \xi_k(r)b_k^\dagger(r)|0\rangle dr=\int_{-\infty}^\infty b^\dag(t)\xi(r)|0\rangle dr
\end{equation}
with $\sum_{k=1}^K\int_{-\infty}^\infty |\xi_k(r)|^2dr=1$ and $\xi(r)=[\xi_1(r)\ \cdot\cdot\cdot\ \xi_K(r)]^T$. Here $|0\rangle$ denotes a multi-channel vacuum input.

The following result will be used later.
\begin{lem}(\cite[Lemma 2]{Guofeng14})\label{lemsi}
Suppose $A$ is Hurwitz. Letting $t_0\rightarrow-\infty$, Eq. (\ref{outlin}) becomes a convolution
\begin{equation}
\breve{b}_{out}(t)=\int_{-\infty}^\infty g_{G}(t-r)\breve{b}(r)dr,
\end{equation}
or equivalently,
\begin{equation}\label{silin}
\breve{b}(t)=\int_{-\infty}^\infty g_{G}(t-r)\breve{f}(r,-\infty)dr.
\end{equation}
Moreover, the stable inversion of Eq. (\ref{silin}) exists and is given by
\begin{equation}
\breve{f}(t,-\infty)=\int_{-\infty}^\infty g_{G^{-1}}(t-r)\breve{b}(r)dr.
\end{equation}
The formula to compute the stable inverse function $g_{G^{-1}}(\cdot)$ is given in \cite[Eq. (19)]{Guofeng13}.
\end{lem}
Following a similar argument as in the proof of Proposition 2 in \cite{Guofeng13}, we have
\begin{lem}\label{leminf}
If the single photon input $|1_\xi\rangle$ is defined by (\ref{sinput}), the output state of the total system in the limit $(t_0\rightarrow-\infty, t\rightarrow\infty)$ can be written as
\begin{equation}\label{fl1}
\rho_\infty=\int_{-\infty}^\infty dr\xi(r)f^\dagger(r,-\infty)\rho_{\infty g}\int_{-\infty}^\infty dr\xi^*(r)f(r,-\infty)
\end{equation}
with $\rho_{\infty g}=\lim_{t\rightarrow\infty,t_0\rightarrow-\infty}U(t,t_0)\rho_0U^\dagger(t,t_0)$, where $\rho_0$ is the initial state of the total system defined as $\rho_0=|0_s\rangle\langle0_s|\otimes|0\rangle\langle0|$.
\end{lem}

\subsection{Quantum finite-level systems}\label{subsec:finite_level_single_photon}

A quantum $N$-level system has states residing in the Hilbert space  $\mathcal H=\mathbb{C}^N$. Let $|0_s\rangle$ denote the ground state of an $N$-dimensional system, and $\{|j_s\rangle,j=1,2,...,N-1\}$ denote its excited basis states. Then $\langle0_s|j_s\rangle=0$ for $j=1,2,...,N-1$. The Pauli operator $\sigma_z$ for a qubit is defined as $\sigma_z=|1_s\rangle\langle1_s|-|0_s\rangle\langle0_s|$. The raising and lowering operators for the qubit are given by $\sigma_+=|1_s\rangle\langle0_s|,\sigma_-=|0_s\rangle\langle1_s|$ respectively.

For a quantum $N$-level system, the commutators $[X,L]$ and $[L^\dag,X]$ in Eq. (\ref{nopr2}) usually are not constant, instead they are often operators, so the term $b^\dagger(t)S^\dagger[X(t),L(t)]+[L^\dagger(t),X(t)]Sb(t)$ in Eq. (\ref{nopr2}) is in general nonlinear in $b(t)$. For example, consider a two-level system described by the triplet $(S,L,H_0)=\left(1,\sqrt{\kappa}\sigma_-,\frac{\omega_c}{2}\sigma_z\right)$. Here, $\omega_c$ is the transition frequency between the ground and excited states.  And $\kappa$ is a parameter defined by $\kappa=2\pi g^2$, where $g$ is the coupling strength between the system and field. For this system, Eq. (\ref{nopr2})  becomes
\begin{eqnarray}
\dot{\sigma}_-(t) &=& -(i\omega_c + \frac{\kappa}{2})\sigma_-(t) +\sqrt{\kappa} \sigma_z (t) b(t), \label{example}\\
b_{out}(t)&=& \sqrt{\kappa} \sigma_-(t) + b(t). \label{example_2}
\end{eqnarray}
Due to nonlinearity,  most studies of the interaction of quantum finite-level systems and single-photon states are often based on various assumptions such as weak excitation limit and are primarily focused on two-level systems.

\section{Main results}\label{secfl}
In this section, we prove that the response of a class of quantum finite-level systems to single-photon input can be analytically solved using a transfer function approach. Recall that $L=\theta^TL_0$.
\begin{lem}\label{lemout2}
Assume the interaction between the plant and the input field is given by the triplet $(S,L,H_0)$ and
\begin{equation} \label{eq:H_L}
H_0|0_s\rangle=\alpha|0_s\rangle,\ L_0|0_s\rangle=0
\end{equation}
 for some constant $\alpha$. Then the following equalities
\begin{equation}\label{lemout2e1}
U(t,t_0)|0\rangle|0_s\rangle=\exp(\mbox{i}\alpha_t)|0\rangle|0_s\rangle,
\end{equation}
and
\begin{equation}\label{lemout2e2}
U^\dagger(t,t_0)|0\rangle|0_s\rangle=\exp(-\mbox{i}\alpha_t)|0\rangle|0_s\rangle,
\end{equation}
hold for some phase shift $\exp(\mbox{i}\alpha_t)$.
\end{lem}
\begin{pf*}{Proof. }
Considering Eq. (\ref{lem1p}) and using the assumptions, we have
\begin{eqnarray}
&&\langle0|\langle0_s|dU(t,t_0)\nonumber\\
&=&\langle0|\langle0_s|\{b^\dag(t)L-L^\dag Sb(t)-(\frac{1}{2}L^\dag L+\mbox{i}H_0)dt\}U(t,t_0)\nonumber\\
&=&-\mbox{i}\alpha dt\langle0|\langle0_s|U(t,t_0).
\end{eqnarray}
Therefore we have $\langle0|\langle0_s|U(t,t_0)=e^{-\mbox{i}\alpha(t-t_0)}\langle0|\langle0_s|$, which proves Eq.~(\ref{lemout2e1}). Eq.~(\ref{lemout2e2}) is obtained using (\ref{lemout2e1}):
\begin{eqnarray}
&&U^\dagger(t,t_0)U(t,t_0)|0\rangle|0_s\rangle=U^\dagger(t,t_0)\exp(\mbox{i}\alpha_t)|0\rangle|0_s\rangle\nonumber\\
&\Rightarrow&U^\dagger(t,t_0)|0\rangle|0_s\rangle=\exp(-\mbox{i}\alpha_t)|0\rangle|0_s\rangle.
\end{eqnarray}\qed
\end{pf*}
Lemma \ref{lemout2} says if the plant Hamiltonian and plant-field coupling don't generate photons, then the field will remain vacuum and the plant  at its ground state.
\begin{lem}\label{lemmanew}
If the input to the $N$-level system is  the single photon input $|1_\xi\rangle$ defined in (\ref{sinput1}), then the output field state is
\begin{eqnarray}\label{fl3}
\rho_{\infty,field}&=&\text{Tr}_s(\rho_\infty)\nonumber\\
&=&\sum_{j=0}^{N-1}\langle j_s|\int_{-\infty}^\infty dtf^\dagger(t,-\infty)\xi(t)|0_s\rangle|0\rangle\nonumber\\
&&\times\langle0|\langle0_s|\int_{-\infty}^\infty dr\xi^\dag(t)f(t,-\infty)|j_s\rangle.
\end{eqnarray}
\end{lem}
\begin{pf*}{Proof. }
Following the techniques in \cite{Guofeng13}, $\rho_\infty$ can be obtained by extending upon the derivation of Lemma \ref{leminf}. Then $\rho_{\infty,field}$ is obtained by tracing out the plant.\qed
\end{pf*}


The following theorem is the main result of this paper, it gives the analytic expression of the output field state of a large class of quantum finite-level systems driven by single-photon input states.
\begin{thm}\label{theorem3}
In addition to the assumptions in Lemma \ref{lemout2}, if
\begin{equation}
\langle0_s|[L_0,H_0]=\langle0_s|\beta L_0,\quad [L_0^\dagger,L_0]|0_s\rangle=h|0_s\rangle\label{fl9},
\end{equation}
and
\begin{equation}\label{fn3}
\Re(a=-\mbox{i}\beta+\frac{1}{2}\sum_{k=1}^K|c_k|^2h)<0,
\end{equation}
hold with $\beta$ and $h$ being constants, then the output field state in response to a single photon input $|1_\xi\rangle$ as defined in (\ref{sinput1}) is given by
\begin{equation}\label{flt4}
\rho_{\infty,field} = |1_{\xi^\prime}\rangle \langle 1_{\xi^\prime}|,
\end{equation}
where the pulse shape $\xi^\prime$ is given by the linear transfer
\[
\xi^\prime (t) =  \int_{-\infty}^\infty g_{G^{-}}(t-r)\xi(r)dr
\]
with the impulse response function being
\begin{equation}
g_{G^{-}}(t) :=\left\{
\begin{array}{ll}
h\theta^T(\theta^T)^\dag e^{at}S+\delta(t)S, &
t\geq 0, \\
0, & t<0.
\end{array}
\right.  \label{eq:tf}
\end{equation}
\end{thm}
\begin{pf*}{Proof. }
$h$ is real because it is an eigenvalue of a Hermitian operator $[L_0^\dagger,L_0]$. According to Lemma \ref{lemmanew}, we need to solve for $\langle0|\langle0_s|f(t,-\infty)|j_s\rangle$. By Eq. (\ref{nopr2}), we have
\begin{eqnarray}
\dot{L_0}(t)&=&\mathcal G_t(L_0)+b^\dagger(t)S^\dag[L_0(t),\theta^TL_0(t)]\nonumber\\
&+&[(\theta^TL_0(t))^\dag,L_0(t)]Sb(t),\label{fl4}\\
b_{out}(t)&=&L(t)+Sb(t).\label{fl5}
\end{eqnarray}
Particularly by (\ref{fl4}) we have
\begin{eqnarray}\label{fl6}
&&\langle0|\langle0_s|\dot{L_0}(t)\nonumber\\
&=&\langle0|\langle0_s|(\mathcal G_t(L_0)+[(\theta^TL_0(t))^\dag,L_0(t)]Sb(t))\nonumber\\
&=&\langle0|\langle0_s|aL_0(t)\nonumber\\
&&+\langle0|\langle0_s|(\theta^T)^\dag U^\dagger(t,t_0)[L_0^\dag,L_0]U(t,t_0)Sb(t)\nonumber\\
&=&\langle0|\langle0_s|(aL_0(t)+h(\theta^T)^\dag Sb(t)),
\end{eqnarray}
where Eq. (\ref{fl9}) and Lemma \ref{lemout2} are used to derive the last line. This is where the single-photon hypothesis plays a key role. It can be seen clearly from Eq. (\ref{fl6}) that the nonlinear stochastic differential equation (\ref{fl4}) is transformed to a linear version under single photon driving. Letting $t_0\rightarrow-\infty$, we can solve Eq. (\ref{fl6}) and then combine with Eq. (\ref{fl5}) to yield
\begin{eqnarray}\label{invb}
&&\langle0|\langle0_s|b_{out}(t)\nonumber\\
&=&\langle0|\langle0_s|\int_{-\infty}^t[h\theta^T(\theta^T)^\dag e^{a(t-r)}+\delta(t-r)]Sb(r)dr.
\end{eqnarray}
By Lemma \ref{lemout2} and $b_{out}(t)=U^\dagger(t,t_0)b(t)U(t,t_0)$, we can express Eq. (\ref{invb}) as
\begin{equation}\label{befsi}
\langle 0|\langle 0_{s}|b(t)=\int_{-\infty }^{\infty }g_{G^{-}}(t-r)\langle 0|\langle0_{s}|f(r,-\infty )dr,
\end{equation}%
where $g_{G^-}$ is that defined in Eq. (\ref{eq:tf}).
Then by Lemma \ref{lemsi} we can apply stable inversion on Eq. (\ref{befsi}) to get
\begin{equation}
\langle 0|\langle 0_{s}|f(t,-\infty )=\left[
\begin{array}{cc}
1 & 0%
\end{array}%
\right] \int_{-\infty }^{\infty }g_{G^{-1}}(t-r)\langle 0|\langle 0_{s}|\breve{b}(r)dr.
\label{eq:sept1_2}
\end{equation}%
Employing Lemma $1$ in \cite{Guofeng13}, we can obtain
\begin{eqnarray*}
&&\left[
\begin{array}{cc}
1 & 0%
\end{array}%
\right] \int_{-\infty }^{\infty }g_{G^{-1}}(t-r)\breve{b}(r)dr
\\
&=&\left[
\begin{array}{cc}
1 & 0%
\end{array}%
\right] \int_{-\infty }^{\infty }\left[
\begin{array}{cc}
g_{G^{-}}(r-t)^{\dag } & 0 \\
0 & g_{G^{-}}(r-t)%
\end{array}%
\right] \left[
\begin{array}{c}
b(r) \\
b^{\dagger}(r)%
\end{array}%
\right] dr  \notag \\
&=&\int_{-\infty }^{\infty }g_{G^{-}}(r-t)^{\dag }b(r)dr.  \notag
\end{eqnarray*}
Eq. (\ref{eq:sept1_2}) becomes%
\begin{equation}
\langle 0|\langle 0_{s}| f(t,-\infty )=\int_{-\infty }^{\infty }g_{G^{-}}(r-t)^{\dag}\langle 0|\langle 0_{s}| b(r)dr.
\end{equation}%
Therefore, the only nonzero term in $\{\langle0|\langle0_s|f(t,-\infty)|j\rangle_s$, $j=0,...,N-1\}$ is $\langle0|\langle0_s|f(t,-\infty)|0\rangle_s$. Then it is straightforward to verify Eq. (\ref{flt4}) is the output field state.\qed
\end{pf*}
It is worth mentioning that the coupling between a two-level system and the input field is commonly modelled by the operator $L_0=\sigma_-$, even for multiple channels. This $L_0$ satisfies the condition Eq. (\ref{fl9}) since $[\sigma_+,\sigma_-]|0_s\rangle=-\frac{I-\sigma_z}{2}|0_s\rangle=-|0_s\rangle$. Moreover, the conditions for $H_0$ are often met. Therefore, as shown in the next section, Theorem \ref{theorem3} is applicable to a wide range of qubit systems. If the coupling is modelled by other system operators, e.g. $L_0=\sigma_x=|0_s\rangle\langle1_s|+|1_s\rangle\langle0_s|$, then the system-environment interaction $\sigma_x(b(t)+b^\dag(t))$ may generate photons when acting on the vacuum state $|0\rangle|0_s\rangle$. When there exist more than one photons, the system may not follow linear dynamics due to nonlinear photon-photon interaction.

Also note that the coupling operator for a linear optical cavity is usually $a_-$ and it satisfies $[a_+,a_-]=-1$, where $a_+$ and $a_-$ are the creation and annihilation operators of the cavity mode respectively.  This explains why the two-level system with $L_0=\sigma_-$ may exhibit linear input-output relation under single photon driving.

\section{Applications}\label{secapp}
In this section three applications drawn from the quantum physics literature are used to illustrate the usefulness of Theorem \ref{theorem3}.

\subsection{Two-level system:one input channel}\label{sectwoone}
We consider the quantum two-level system  (\ref{example})-(\ref{example_2}) driven by a single-photon state (\ref{sinput}). Applying Theorem \ref{theorem3}, the output pulse shape is calculated to be
\begin{equation}
\xi^{'}(t)=\int_{-\infty }^{t}[-\kappa e^{-\left( \frac{\kappa }{2}+\mbox{i}\omega _{c}\right) (t-r)}+\delta (t-r)]\xi (r)dr.  \label{eq:eta}
\end{equation}
Also, we can easily obtain the Fourier transform of (\ref{eq:eta}) using the convolution theorem:
\begin{equation}\label{outputp2}
\tilde{\xi}^{'}(\omega)=\tilde{\xi}(\omega)\frac{-\frac{\kappa}{2}+\mbox{i}(\omega+\omega_c)}{\frac{\kappa}{2}+\mbox{i}(\omega+\omega_c)} :=\tilde{\xi}(\omega)G(\mbox{i}\omega),
\end{equation}
The single photon response of two-level systems has been extensively studied in physics, see e. g. \cite{Shen:05,Zhou08}. Here we obtained the analytic form of the output field state without making any physical approximations such as weak excitation limit or scattering modes. Compared to the results obtained in \cite{Guofeng13}, the output state is analogous to the output of a single-mode linear system in response to a single-photon input. This observation is consistent with the existing results from \cite{Shen:05,Fan10,Zhou08}, where the authors have found that the transmission and reflection spectrums for the single-photon transport through a two-level system are analogous to the scattering spectrums for linear cavities. We can apply zero-dynamics principle \cite{Naoki14} for studying the full inversion of the states. To remove the zero from the transfer function, we should choose the input as $\tilde{\xi}(\mbox{i}\omega)=\sqrt{\kappa}/(-\frac{\kappa}{2}+\mbox{i}\omega+\mbox{i}\omega_c)$. The inverse Fourier transform of the input yields $\xi(t)=-\sqrt{\kappa}e^{(\frac{\kappa }{2}-\mbox{i}\omega _{c})t}(1-u(t))$, with $u(t)$ being the Heaviside step function. The input pulse is exponentially rising but with a resonant phase component till $t=0$ in time domain. This inverting single-photon pulse matches the existing designs for the inversion of two-level atoms \cite{Cirac97,Rephaeli10}.

\subsection{Two-level system:two input channels}
The system is described by the triplet $(S,L,H_0)=(I_{2},[\sqrt{\kappa _{1}}\ \sqrt{\kappa _{2}}]^T\sigma _{-},\frac{\omega _{c}}{2}\sigma _{z})$.
Let the input photon enter the plant from the first channel
\begin{equation}\label{ttwo22}
|1_{\xi_1}\rangle=\int_{-\infty}^\infty \xi_1(r)b_1^\dagger(r)|0\rangle dr.
\end{equation}
In this case, the second-channel input is a vacuum state. The output field state is calculated to be $|\Psi _{out}\rangle\langle\Psi _{out}|$, where we define
\begin{equation}
\left\vert \Psi _{out}\right\rangle=\int_{-\infty }^{\infty }\xi_1^{'}(t)b_1^{\dagger}(t)dt|0\rangle+\int_{-\infty }^{\infty }\xi_2^{'}(t)b_2^{\dagger}(t)dt|0\rangle.
\end{equation}
The shapes of the output pulses in these two channels are given by $\xi_1^{'}(t)=\xi_1(t)-\kappa _{1}\eta (t)$ and $\xi_2^{'}(t)=-\sqrt{\kappa _{1}\kappa _{2}}\eta (t)$, and $\eta(t)$ is expressed as
\begin{equation}
\eta (t):=\int_{-\infty }^{t}\ \ e^{-\left( i\omega _{c}+\frac{\kappa
_{1}+\kappa _{2}}{2}\right) (t-r)}\xi_1(r)dr.  \label{eta}
\end{equation}
Define two transfer functions $G_1(\mbox{i}\omega)$ and $G_2(\mbox{i}\omega)$ in terms of $\tilde{\xi}_1^{'}(\omega)=G_1(\mbox{i}\omega)\tilde{\xi_1}(\omega)$ and $\tilde{\xi}_2^{'}(\omega)=-G_2(\mbox{i}\omega)\tilde{\xi_1}(\omega)$.
Simple calculation yields
\begin{eqnarray}
G_1(\mbox{i}\omega)=\frac{-\frac{\kappa_1-\kappa_2}{2}+\mbox{i}(\omega+\omega_c)}{\frac{\kappa_1+\kappa_2}{2}+\mbox{i}(\omega+\omega_c)},\label{fs23}\\
G_2(\mbox{i}\omega)=\frac{\sqrt{\kappa_1\kappa_2}}{\frac{\kappa_1+\kappa_2}{2}+\mbox{i}(\omega+\omega_c)}.\label{fs24}
\end{eqnarray}
Physically, $G_1(\mbox{i}\omega)$ could correspond to a transmission spectrum and $G_2(\mbox{i}\omega)$ could be related to reflection.  Further calculation shows
\begin{equation}\label{sb2}
|G_1(\mbox{i}\omega)|^2=1-\frac{4\kappa_1\kappa_2}{(\kappa_1+\kappa_2)^2+4(\omega+\omega_c)^2}.
\end{equation}
By (\ref{sb2}), if $\omega$ is largely detuned from the transition frequency $\omega_c$ of the qubit, we have $|G_1(\mbox{i}\omega)|^2\approx1$ and the photon will transmit with high probability. On the other hand,
\begin{equation}\label{sb4}
|G_2(\mbox{i}\omega)|^2=\frac{4\kappa_1\kappa_2}{(\kappa_1+\kappa_2)^2+4(\omega+\omega_c)^2}.
\end{equation}
Since $(\kappa_1+\kappa_2)^2\geq4\kappa_1\kappa_2$ is always true, $|G_2(\mbox{i}\omega)|^2=1$ admits a solution if and only if $\kappa_1=\kappa_2$ and $\omega=-\omega_c$. Only the frequency component in resonance with the qubit can be perfectly reflected if the two channels are coupled to the qubit with the same strength. If $\kappa_1\neq\kappa_2$, there is no frequency component that can be perfectly reflected. These results support a theoretical understanding of the numerical studies in \cite{Bara12}.

\subsection{Gradient echo quantum memory}
Consider the finite input-output model for gradient echo memories where $N$ two-level atoms are interconnected by series product \cite{Gough09} via one channel \cite{Hush13} . The $(S,L,H_0)$ representation of the system is given by $(I,\sum_{n=1}^N\sqrt{\kappa}\sigma_-^n,\sum_{n=1}^N\frac{\omega_c}{2}\sigma_z^n+\frac{\kappa}{2\mbox{i}}\sum_{j=2}^N\sum_{i=1}^{j-1}(\sigma_+^j\sigma_-^i-\sigma_+^i\sigma_-^j))$. A so-called weak atomic excitation limit is introduced in \cite{Protsenko99,Hush13} to approximate the atoms by linear cavities in this memory model.

When the input to the memory is a single-photon state, then the output of each atom is a single-photon state as well, which implies that each two-level atom of the memory is driven by a single-photon state. As a consequence, we can simply use the result from Section~\ref{sectwoone} to obtain the pulse function of the output state of the memory as
\begin{equation}\label{reviseeq1}
\tilde{\xi}^{'}(\omega)=(\frac{-\frac{\kappa}{2}+\mbox{i}(\omega+\omega_c)}{\frac{\kappa}{2}+\mbox{i}(\omega+\omega_c)})^N\tilde{\xi}(\omega).
\end{equation}
The inverse Fourier transform of (\ref{reviseeq1}) yields the time-domain output field state
\begin{eqnarray}
\xi^{'}(t)&=&\int_{-\infty}^{t}dr[\kappa N{e^{-\frac{\kappa}{2}(t-r)}}_1F_1(1+N,2,-\kappa(t-r))\nonumber\\
&+&\delta(t-r)]\xi(r),
\end{eqnarray}
where we have used the the definition of Kummer confluent hypergeometric function $_1F_1(a,b,z)=\sum_{k=0}^{\infty}\frac{a^{(n)}z^n}{b^{(n)}n!}$ with $a^{(n)}=a(a+1)\cdot\cdot\cdot(a+n-1)$ \cite{abramowitz1964handbook}. Note that the above procedure can be easily extended to calculate the output state for atoms of different resonant frequencies $\{\omega_c^n\}$.

\begin{figure}
\includegraphics[scale=0.9]{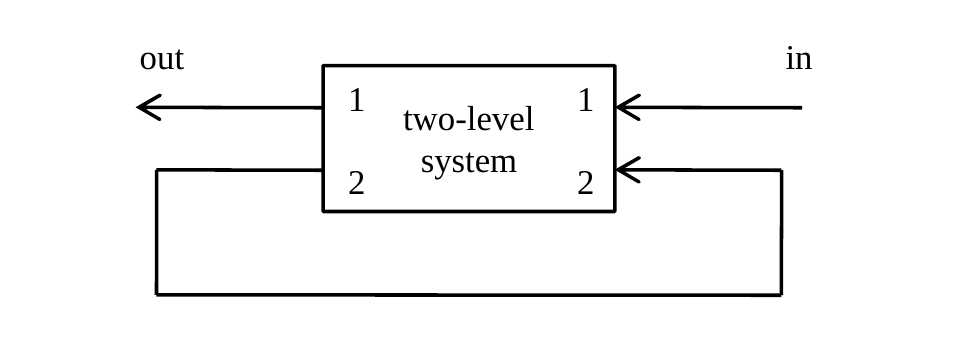}
\caption{The second output channel is directly fed back to the system as the second input channel. The entry $S_{ij}$ in $S$ is the scattering coefficient from the $i$-th input channel to the $j$-th output channel.}
\label{figonemode}
\end{figure}

\section{Single-photon pulse shaping by coherent feedback}\label{sec:control}
In this section we study how to manipulate the pulse shape of single-photon states by means of coherent feedback. The idea behind the coherent feedback is that only quantum systems are used for control and hence all the components have similar characteristic time scales. For example, all-optical feedback has been used in practical control design which removes the slow and inefficient measurement process from the feedback loop. For experimental demonstrations of coherent feedback please see \cite{PhysRevA.78.032323,Iida12}.

Consider a two-channel two-level system with parameters
\[
(S,[\sqrt{\kappa _{1}}\ \sqrt{\kappa _{2}}]^T\sigma _{-},\frac{\omega _{c}}{2}\sigma _{z}).
\]
We design a coherent feedback by linear fractional transformation, as shown in Figure \ref{figonemode}. In what follows we study two cases.

Case 1:  $S$ is real. The resulting single-channel  system is given by the triplet $(S_{11}+S_{12}(1-S_{22})^{-1}S_{21},(\sqrt{\kappa_1}+S_{12}(1-S_{22})^{-1}\sqrt{\kappa_2})\sigma_-,\frac{\omega _{c}}{2}\sigma _{z})$. Using Theorem \ref{theorem3}, we can solve for the transfer function of this system to be
\begin{eqnarray}
&&G(\mbox{i}\omega)\nonumber\\
&=&(S_{11}+\frac{S_{12}S_{21}}{1-S_{22}})\frac{-\frac{(\sqrt{\kappa_1}+\frac{S_{12}}{1-S_{22}}\sqrt{\kappa_2})^2}{2}+\mbox{i}(\omega+\omega_c)}{\frac{(\sqrt{\kappa_1}+\frac{S_{12}}{1-S_{22}}\sqrt{\kappa_2})^2}{2}+\mbox{i}(\omega+\omega_c)}.\nonumber\\
\end{eqnarray}
This suggests we can use $S$ and $\kappa_2$ to control the input-output transfer function, and thus shape the output pulse. For example, if $S=[0\ 1;1\ 0]$, i. e. the first-channel input is scattering to the second-channel and directly fed back to the system, then $G(\mbox{i}\omega)=(-\frac{(\sqrt{\kappa_1}+\sqrt{\kappa_2})^2}{2}+\mbox{i}(\omega+\omega_c))/(\frac{(\sqrt{\kappa_1}+\sqrt{\kappa_2})^2}{2}+\mbox{i}(\omega+\omega_c))$. The decay of the two-level system is enhanced and the transfer spectrum is made sharper compared to that in Eq. (\ref{outputp2}).

Case 2: $S$ is complex-valued. A complex-valued matrix $S$ can be realized by interconnecting a beam-splitter before the inputs entering the plant. In this case, the triplet of the feedback network is given by $(S_{11}+S_{12}(1-S_{22})^{-1}S_{21},(\sqrt{\kappa_1}+S_{12}(1-S_{22})^{-1}\sqrt{\kappa_2})\sigma_-,\frac{\omega _{c}}{2}\sigma _{z}+\frac{\sigma_z+1}{2}\Im(\sqrt{\kappa_1\kappa_2}S_{12}(1-S_{22})^{-1}+\kappa_2S_{22}(1-S_{22})^{-1}))$.  Using Theorem \ref{theorem3} and denoting $\Delta :=\Im(\sqrt{\kappa_1\kappa_2}S_{12}(1-S_{22})^{-1}+\kappa_2S_{22}(1-S_{22})^{-1})$, we can solve for the transfer function to be
\begin{eqnarray}
&&G(\mbox{i}\omega)=(S_{11}+S_{12}S_{21}(1-S_{22})^{-1})\nonumber\\
&&~~~~~~\times\frac{-\frac{(\sqrt{\kappa_1}+\sqrt{\kappa_2}S_{12}(1-S_{22})^{-1})^2}{2}+\mbox{i}(\omega+\omega_c+\Delta)}{\frac{(\sqrt{\kappa_1}+\sqrt{\kappa_2}S_{12}(1-S_{22})^{-1})^2}{2}+\mbox{i}(\omega+\omega_c+\Delta)}.
\end{eqnarray}
Therefore, we can further shift the spectrum by an amount of $\Delta$. For example, if a $50/50$ beam-splitter is used, i.e. $S=\frac{1}{\sqrt{2}}[1\ \mbox{i};\mbox{i}\ 1]$, then $G(\mbox{i}\omega)=-(-\frac{(\sqrt{\kappa_1}+\mbox{i}\sqrt{\kappa_2}/(\sqrt{2}-1))^2}{2}+\mbox{i}(\omega+\omega_c+\sqrt{\kappa_1\kappa_2}/(\sqrt{2}-1)))/(\frac{(\sqrt{\kappa_1}+\mbox{i}\sqrt{\kappa_2}/(\sqrt{2}-1))^2}{2}+\mbox{i}(\omega+\omega_c+\sqrt{\kappa_1\kappa_2}/(\sqrt{2}-1)))$.

\section{Conclusion}\label{seccon}
In this paper the response of a class of quantum finite-level systems to single-photon states has been investigated. Analytic expression of the output single-photon states has been derived. Single-photon pulse shaping by means of coherent feedback has also been studied. The future research would include the application of this work to the hybrid coherent quantum networks driven by single photons.

\begin{ack}                               
This research is supported in part by Australian Research Council,  AFOSR Grant FA2386-12-1-4075, National Natural Science Foundation of China grant (No. 61374057, 11404113) and Hong Kong RGC grant (No. 531213).  
\end{ack}

\bibliographystyle{plain}        
\bibliography{ref}        



\end{document}